\documentclass[doublecol]{epl2} 
\usepackage{nicefrac}

\title{Multi-orbital nature of the spin fluctuations in Sr$_2$RuO$_4$}

\author{Lewin Boehnke,\inst{1} Philipp Werner,\inst{1} and Frank Lechermann\inst{2}}
\shortauthor{Lewin Boehnke, Philipp Werner and Frank Lechermann}

\institute{                    
  \inst{1} Department of Physics, University of Fribourg, 1700 Fribourg, Switzerland\\
  \inst{2} I. Institut f{\"u}r Theoretische Physik, Universit{\"a}t Hamburg, 
D-20355 Hamburg, Germany
}
\pacs{71.27.+a}{First pacs description}
\pacs{75.40.Gb}{Second pacs description}
\pacs{74.70.Pq}{Third pacs description}

\abstract{The spin susceptibility of strongly correlated Sr$_2$RuO$_4$ is
known to display a rich structure in reciprocal space, with a prominent peak at
${\bf Q}_i$=$(0.3,0.3,0)$. It is still debated if the resulting 
incommensurate spin-density-wave fluctuations foster unconventional superconductivity
at low temperature or compete therewith. By means of density functional theory combined
with dynamical mean-field theory, we reveal the realistic multi-orbital signature of the 
(dynamic) spin susceptibility beyond existing weak-coupling approaches. The experimental
fluctuation spectrum up to 80 meV is confirmed by theory. Furthermore, the peak at 
${\bf Q}_i$ is shown to carry nearly equal contributions from each of the 
Ru$(4d)$-$t_\mathrm{2g}$ orbitals, pointing to a cooperative effect resulting in the 
dominant spin fluctuations.}

\begin{document}

\maketitle

\section{Introduction}
The Ruddlesden-Popper series of the strontium ruthenates Sr$_{n+1}$Ru$_n$O$_{3n+1}$,
with $n$ labeling the layers of corner-sharing RuO$_6$ octahedra separated by SrO 
rocksalt layers, poses a formidable correlated-electron problem in a realistic
scenario. Intricate competitions between Fermi-liquid, Mott-critical and 
superconducting behavior are accompanied by a complex metallic magnetism.
While the perovskite SrRuO$_3$ ($n$$\rightarrow$$\infty$) as series end member 
is ferromagnetic (FM) below $T_{\rm C}\sim 165$\,K, a ferromagnetic-to-paramagnetic
transition occurs by lowering the number of layers $n$. This transition takes place between the
orthorhombic $n=3$ and $n=2$ systems. The Sr$_4$Ru$_3$O$_{10}$ compound is verified 
FM~\cite{cra02}, whereas bilayer Sr$_3$Ru$_2$O$_7$ remains paramagnetic (PM) down to 
lowest temperatures. However, the bilayer material is prone to FM 
order~\cite{ike00,beh12} with puzzling metamagnetic (MM) behavior in an applied field below 
$T_{\rm MM}\sim 1$\,K (see Ref.~\cite{mac12} for a review).

The tetragonal $n=1$ compound Sr$_2$RuO$_4$ stands out, since it displays 
unconventional superconductivity (most probably triplet pairing), below 
$T_{\rm c} \sim 1.5$\,K~\cite{mae94,ric95,mac03}. De Haas-van Alphen 
measurements~\cite{mac96} and angle-resolved photoemission 
spectroscopy (ARPES)~\cite{yok96,dam00,she01} documented the quasi-two-dimensional (2D) 
electron structure for this single-layer ruthenate. The strongly correlated nature 
is furthermore proven by a large mass renormalization~\cite{mae97}, originating from 
the cooperation between less-screened Coulomb interactions and the actual filling of the 
$t_\mathrm{2g}$ manifold within the Ru$(4d)$ shell. More specifically, by combining 
band-structure and dynamical mean-field theory (DMFT) it was shown that the intriguing 
interplay between the intra-orbital Hubbard interaction $U$ and the inter-orbital Hund's 
exchange $J_{\rm H}$ gives rise to a so-called Hund's metal~\cite{mra11,wer08}. Though 
paramagnetic above the superconducting temperature, Sr$_2$RuO$_4$ exhibits substantial magnetic
correlations, either interlinked or competing with the pairing instability~\cite{huo13}.
A prominent peak in the $\bf q$-dependent spin susceptibility at finite
${\bf Q}_i$=$(0.3,0.3)$ in the ${\bf q}_z=0$ plane~\cite{maz99,sid99} renders it obvious that
antiferromagnetic(-like) spin fluctuations must still be an essential feature of the puzzling 
ruthenate physics.

In this letter we advance on the understanding of the pecularities of Sr$_2$RuO$_4$
by connecting the strong correlation specification to the spin-fluctuations phenomenology.
From the merger of density functional theory (DFT) with DMFT, we have a realistic many-body
theory at hand that goes beyond existing studies~\cite{maz99,kee00,mor01,ere02} 
within the weak-coupling random-phase-approximation (RPA).
We introduce a means of investigation of fluctuation characteristics by a rigorous 
eigen-analysis of the full orbital dependence of the complete susceptibility tensor. 
From that it is shown that multi-orbital correlations known to be relevant for the one-particle 
spectral function are also vital in the two-particle response functions such as the spin
susceptibility $\chi({\bf q},\omega)$. Namely, albeit the contribution of the 
Ru$(4d)$-$t_\mathrm{2g}$ orbitals with nominal total four-electron filling from the 
Ru$^{4+}$ oxidation state differs in ${\bf q}$-space, they equally take part in the formation of 
the dominant peak at ${\bf Q}_i$. This points to the relevance of including the 
contributions of all three $t_\mathrm{2g}$ orbitals when accounting for spin fluctuations in
Sr$_2$RuO$_4$.

\section{Calculational procedure}
\begin{figure}[t]
\includegraphics*[height=1.85cm]{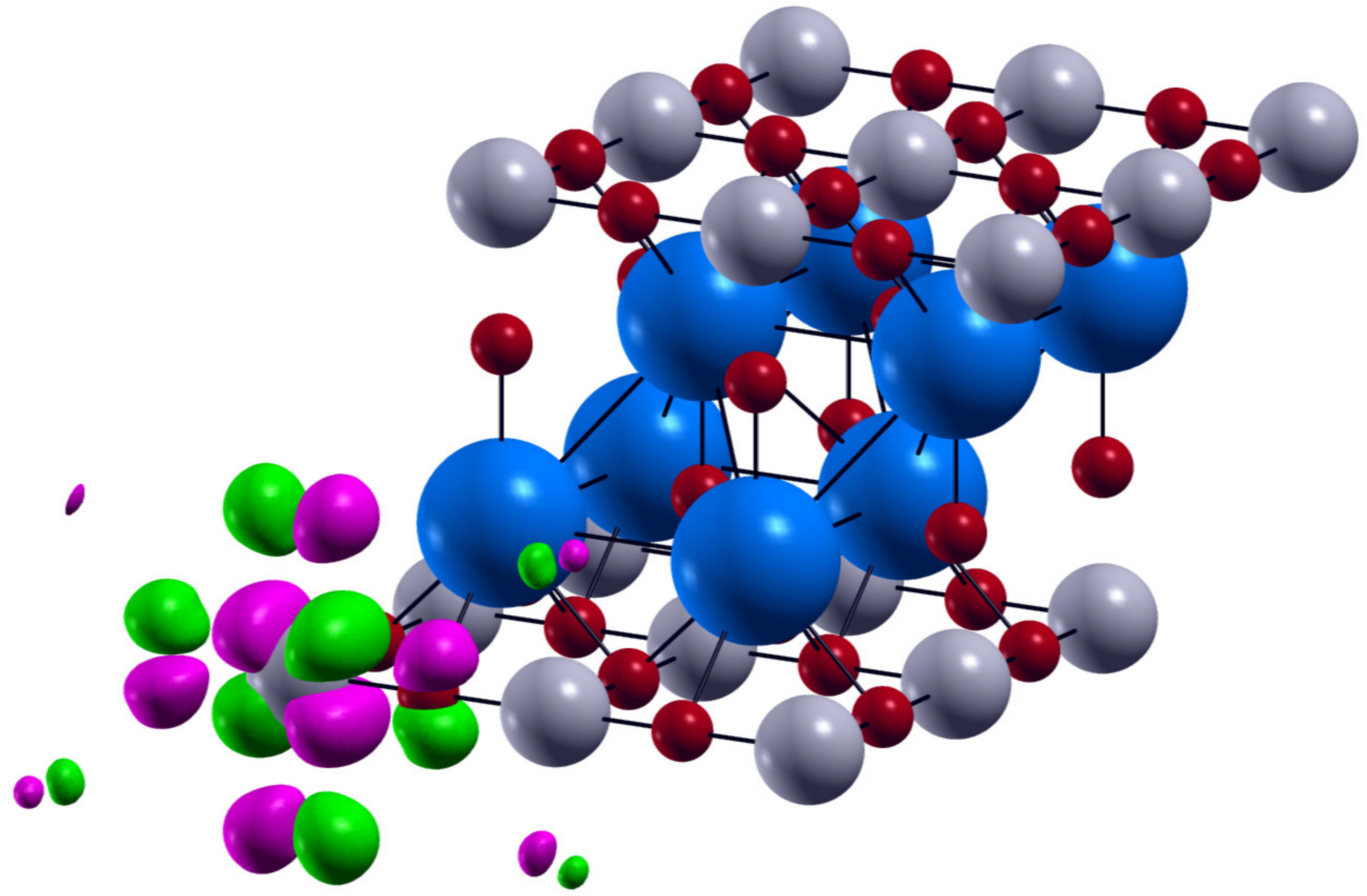}
\includegraphics*[height=1.85cm]{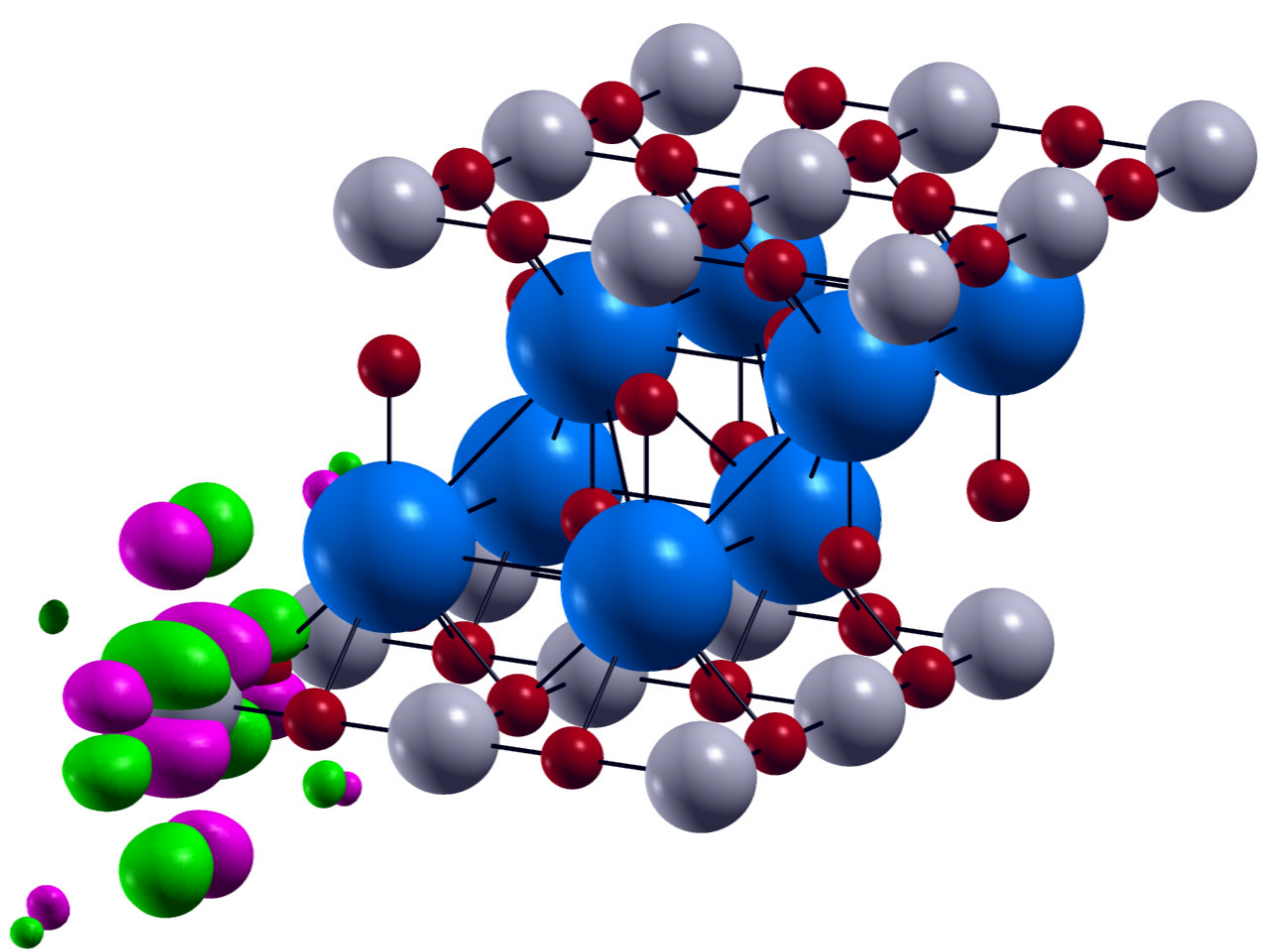}
\includegraphics*[height=1.7cm]{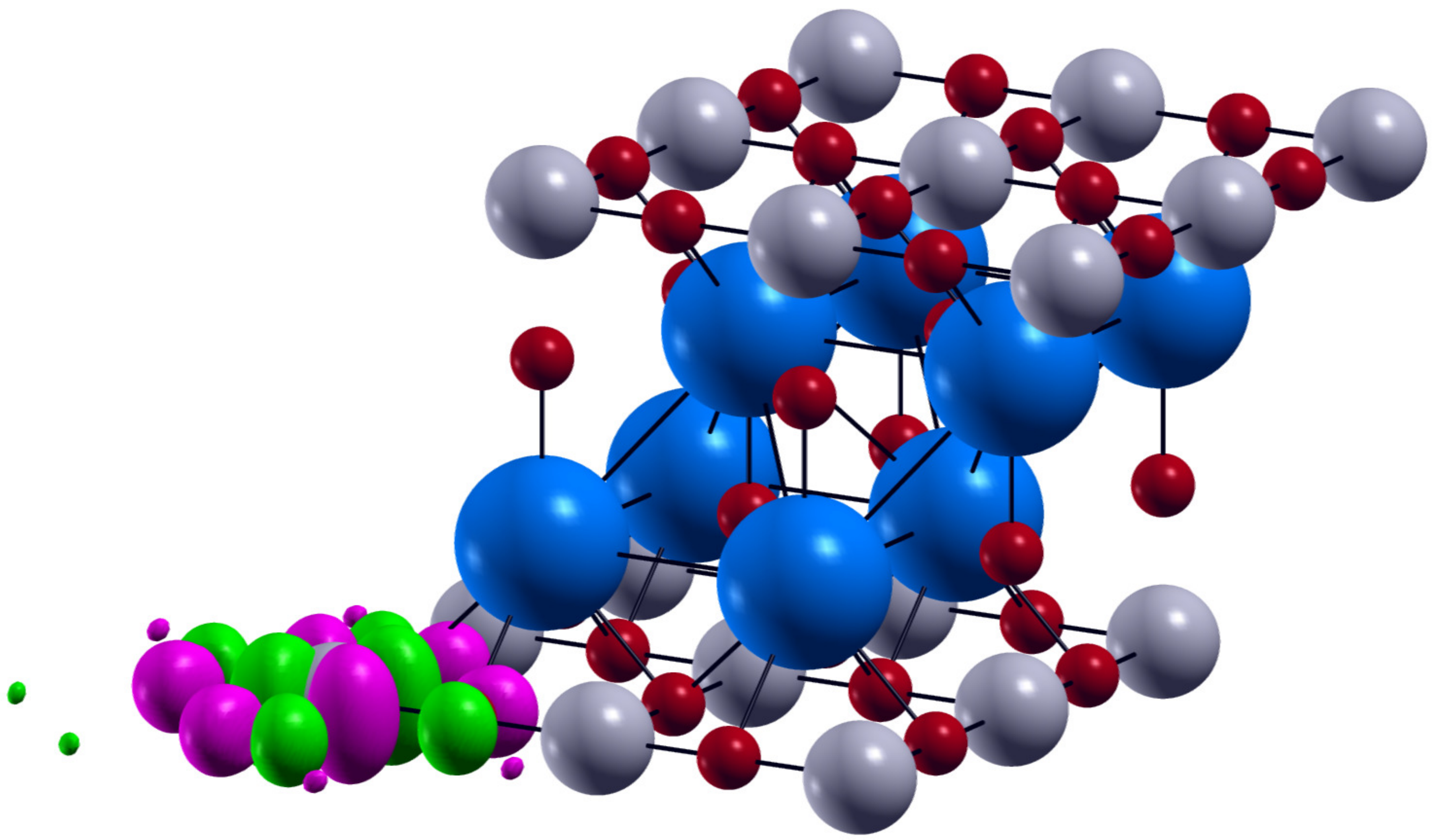}\\
\hspace*{1cm}$d_{xz}$\hspace*{2.2cm}$d_{yz}$\hspace*{2.4cm}$d_{xy}$
\caption{(Color online) Maximally-localized Ru($t_\mathrm{2g}$)-like Wannier orbitals
for Sr$_2$RuO$_4$ in the $I4/mmm$ crystal structure. 
Sr: blue, Ru: grey, O: red.}\label{fig:orbitals}
\end{figure}

Single-layer Sr$_2$RuO$_4$ has ideal tetragonal symmetry (space group $I4/mmm$) with  
fourfold rotation $C_4^z$ around the $c$-axis.
Our first-principles many-body calculations are based on the structural data of Walz and 
Lichtenberg~\cite{wal93}. A mixed-basis pseudopotential scheme is used for the
band-structure part within the local density approximation (LDA)~\cite{mbpp_code}. 
To account for strong electronic correlations in the DMFT treatment, a correlated subspace is 
constructed from the derived Ru $4d(t_\mathrm{2g})$-like maximally localized Wannier 
functions~\cite{mar97,sou01} (see Fig.~\ref{fig:orbitals}). Therein a three-orbital 
Hubbard model in rotationally-invariant Slater-Kanamori paramaterization with $U=2.3$\,eV 
and $J_{\rm H}=0.4$\,eV is constructed~\cite{mra11}. The multi-orbital DMFT impurity problem 
is solved by the continuous-time quantum-Monte Carlo method in the hybridization-expansion 
formulation~\cite{rub05,wer06,boe11,par15,set16}. As in previous 
DFT+DMFT studies~\cite{lie00,mra11}, effects of spin-orbit coupling are not included 
in the present work.

In order to access the two-particle response, we compute the complete three-orbital 
particle-hole susceptibility tensor $\chi^{\sigma\sigma'}_{mm'm''m'''}({\bf q},\omega)$, 
with $m,m'm'',m'''=1,2,3$ and $\sigma,\sigma'=\uparrow,\downarrow$, at finite 
temperature, with full generality in the 
frequency-dependent structure~\cite{boe11}, employing the numerically beneficial orthogonal polynomial basis for the fermionic degrees of freedom. The full ${\bf q}$-dependence is obtained from the DMFT two-particle 
Green's function by inverting the Bethe-Salpeter equation, assuming the locality of the 
irreducible vertex~\cite{zla90,maier05}. For the bosonic dynamic $\omega$-dependence of the susceptibility a further numerically exact optimization was employed, which alters the shift in the fermionic transformation depending on the bosonic frequency \cite{boe15}.

To cope with the multi-orbital character, we focus on the eigenvalues/modes of the 
susceptibility tensor $\chi_{\kappa\kappa'}$ in the superindices $\kappa=\{\sigma mm'\}$ 
and $\kappa'=\{\sigma' m'''m''\}$~\cite{boe14}. Specifically
\begin{eqnarray}
  \label{eqn:eigendecomposition}
  \chi^{(l)}({\bf q},\mbox{$\omega$=0})=&\left<\mathcal{T}_\tau\sum_\kappa 
\hat{v}^{(l)}_\kappa({\bf q}) 
\sum_{\kappa'}\hat{v}^{\star\,(l)}_{\kappa'}({\bf q})\right>\nonumber\\
=&\left<\mathcal{T}_\tau\hat{V}^{(l)}({\bf q})\hat{V}^\star{}^{(l)}({\bf q})\right>
\label{eq:eig}
\end{eqnarray}
is the $l$th eigenvalue with $\hat{V}^{(l)}=\sum_{\sigma mm'}
v^{(l)}_{\sigma mm'}c^\dagger{}^\sigma_mc^\sigma_{m'}$ as the corresponding fluctuating 
eigenmode ($v^{(l)}_\kappa$ is the $l$th eigenvector of $\chi_{\kappa\kappa'}$).

\section{Results}
\begin{figure}[t]
(a)\hspace*{-0.4cm}\includegraphics*[width=8.5cm]{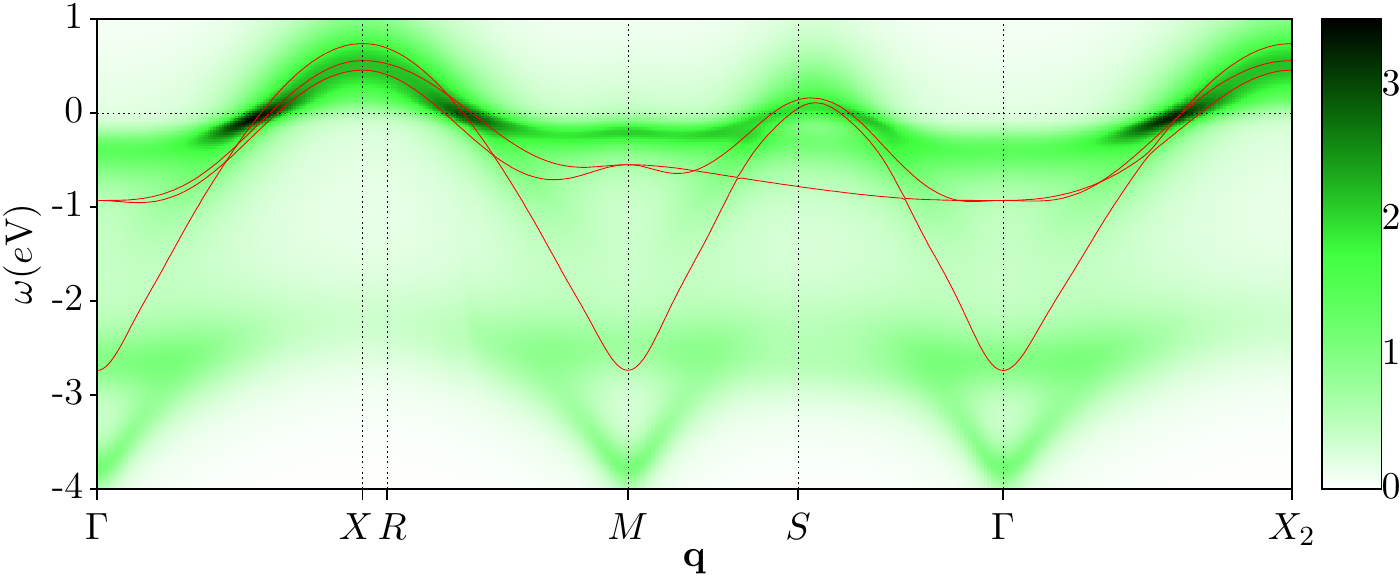}\\
(b)\hspace*{-0.4cm}\includegraphics*[width=4.1cm]{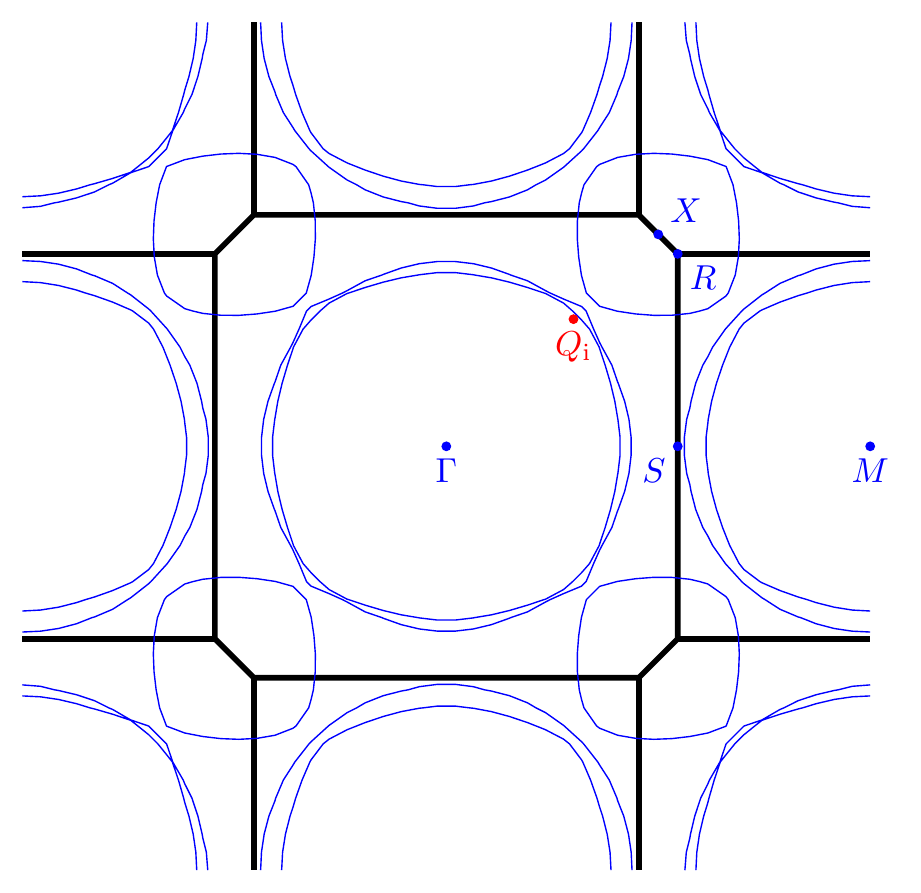}
(c)\hspace*{-0.2cm}\includegraphics*[width=4.1cm]{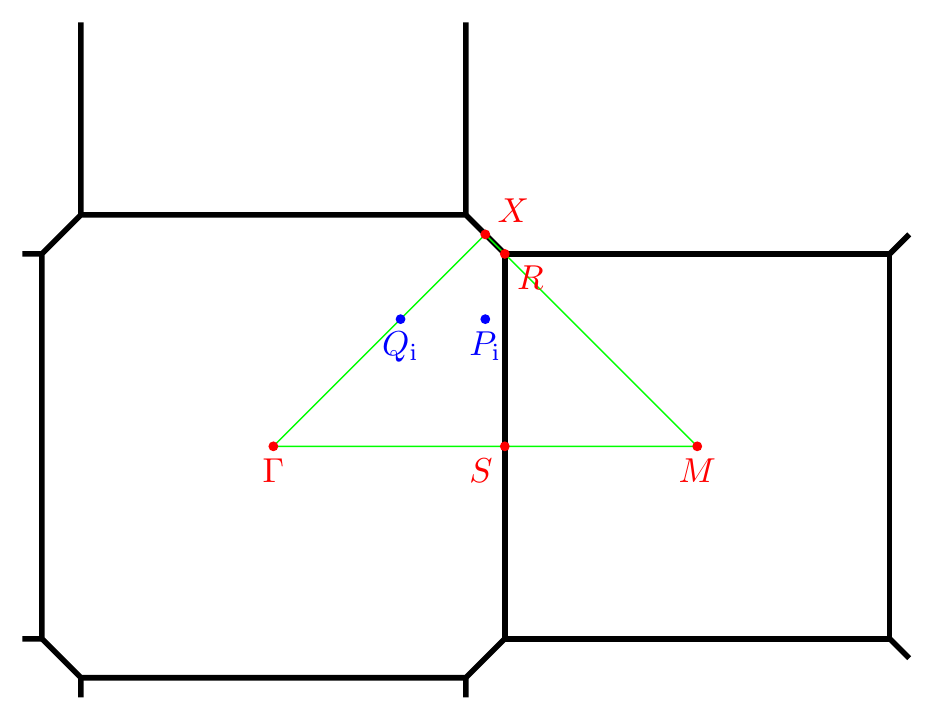}
\caption{(Color online) Correlated electronic structure of Sr$_2$RuO$_4$ from
DFT+DMFT.
(a) one-particle spectral function (full lines: LDA data) and 
(b) Fermi surface in the $k_z$=0-plane.
(c) High-symmetry points in reciprocal space. The $X_2$ point is located above $X$ on
the same level as the upper surface of the Brillouin zone.}\label{fig:one-particle}
\end{figure}

Based on our DFT+DMFT calculations, we first provide some results on the Sr$_2$RuO$_4$ 
correlated electronic structure consistent with previous theoretical work~\cite{lie00,mra11}. 
Figure~\ref{fig:one-particle}a displays the one-particle spectral function
$A({\bf k},\omega)$ along high-symmetry lines in the corresponding Brillouin zone (BZ) (cf. 
Fig.~\ref{fig:one-particle}c). Four electrons reside in the low-energy $t_\mathrm{2g}$-like
manifold with LDA bandwidth $W\sim3.4$\,eV. Electron-electron interactions lead to
significant band renormalization, with additional spectral-weight transfer to $\sim -3$\,eV. 
The wider low-energy band is of dominant $d_{xy}$ character, while the two narrower bands 
are mainly formed by $d_{xz}$, $d_{yz}$. The resulting Fermi surface has three sheets: 
$\alpha$ centered around $X$ and $\beta$ centered around $\Gamma$, both stemming from the 
narrower quasi-onedimensional (quasi-1D) bands, as well as $\gamma$ (also around $\Gamma$) 
derived from the quasi-2D $d_{xy}$-like band (see Fig.~\ref{fig:one-particle}b). 

\begin{figure}[t]
(a)\hspace*{-0.4cm}\includegraphics*[width=8.75cm]{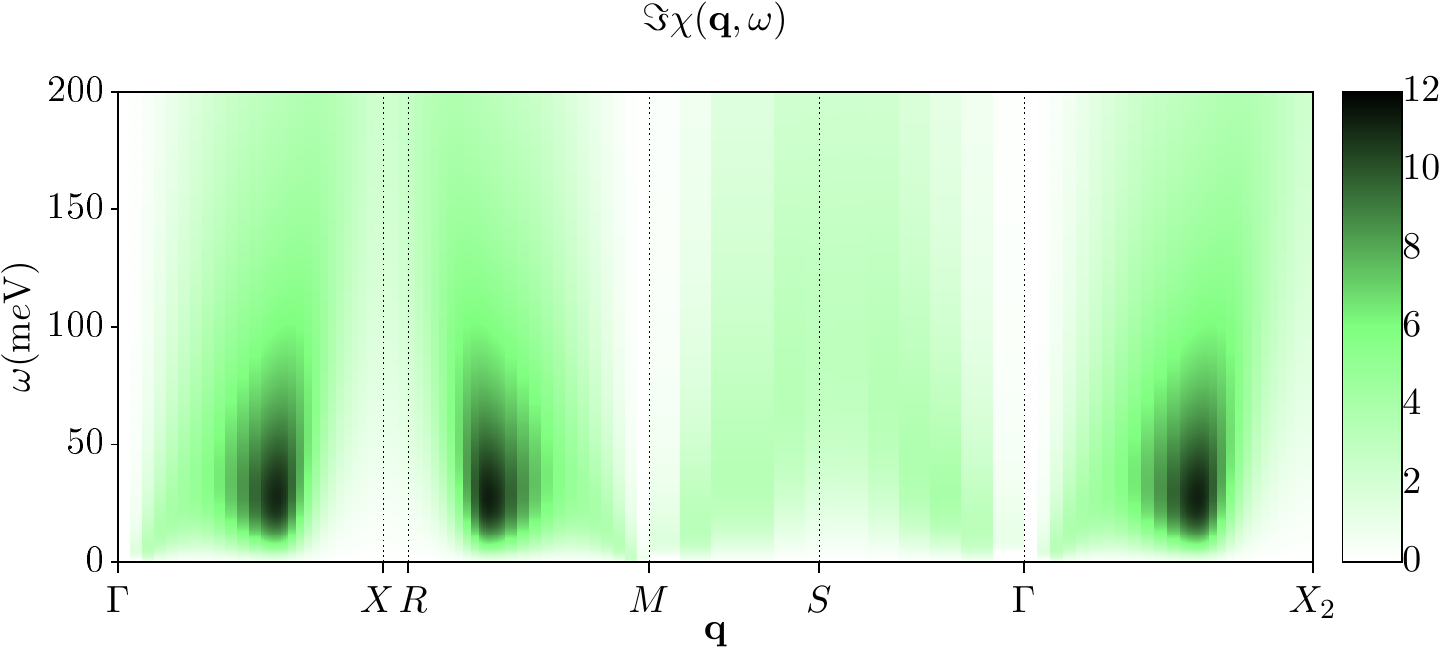}\\
(b)\hspace*{-0.4cm}\includegraphics*[width=8.15cm]{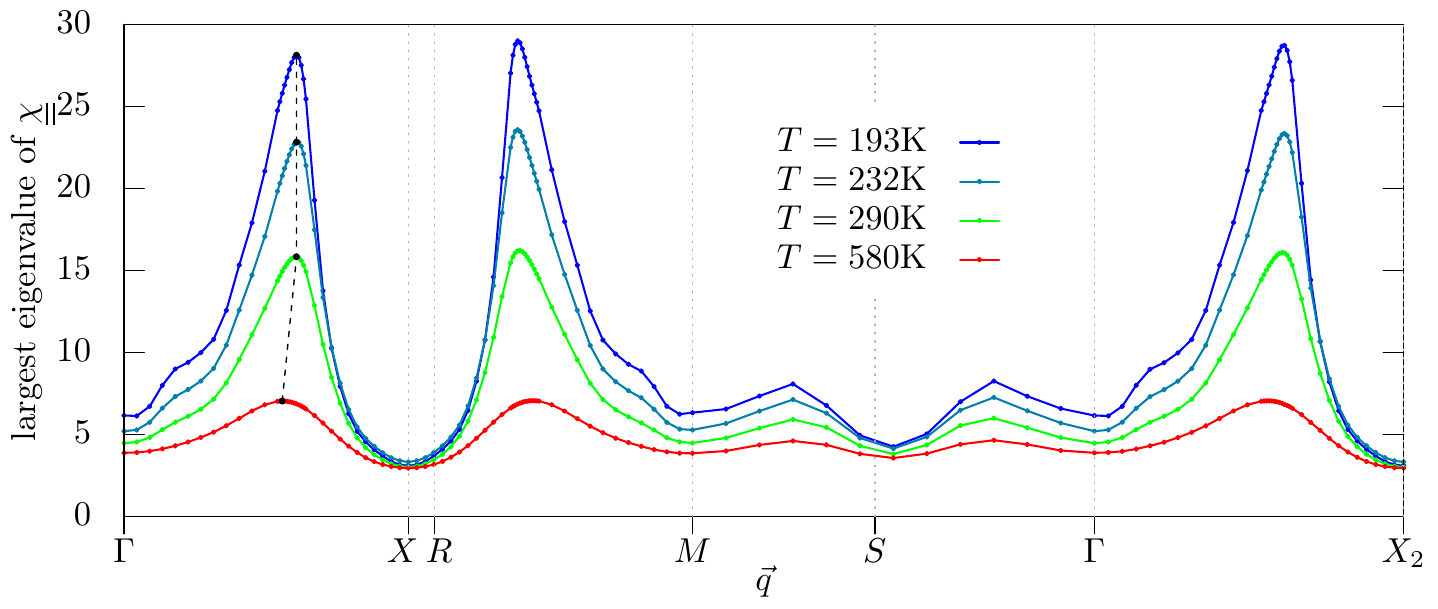}
\caption{(Color online) 
Data on the spin susceptibility along high-symmetry lines in reciprocal space.
(a) Dynamic spin susceptibility $\chi_s({\bf q},\omega)$ at $T=290$\,K.
(b) Dispersion of the dominant eigenvalue of the spin-susceptibility 
tensor for selected temperatures.}\label{fig:sus}
\end{figure}

Due to the possibility of producing very clean samples for single-layer ruthenate, 
inelastic neutron scattering (INS) experiments elucidated the magnetic excitations in
great detail~\cite{sid99,bra02,bra04,iid11}. Though it remains difficult to reach the 
experimental very low-$T$ regime with quantum Monte Carlo, the computed room-temperature 
dynamic spin susceptibility $\chi_s({\bf q},\omega)$ (see Fig.~\ref{fig:sus}a) agrees very 
well with available data. Note that also recent INS reported sizeable fluctuations even up to
that temperature range~\cite{iid11}. The location of the incommensurate peak ${\bf Q}_i$ as 
well as its position in energy $\omega_{\bf Q} \sim 27$\,meV is consistent with 
INS~\cite{bra04,iid11}. Moreover, the experimental finding of a fluctuation regime up to 
80\,meV~\cite{iid11} is confirmed by our calculations.
\begin{figure}[t]
(a)\hspace*{-0.25cm}\includegraphics*[width=8.25cm]{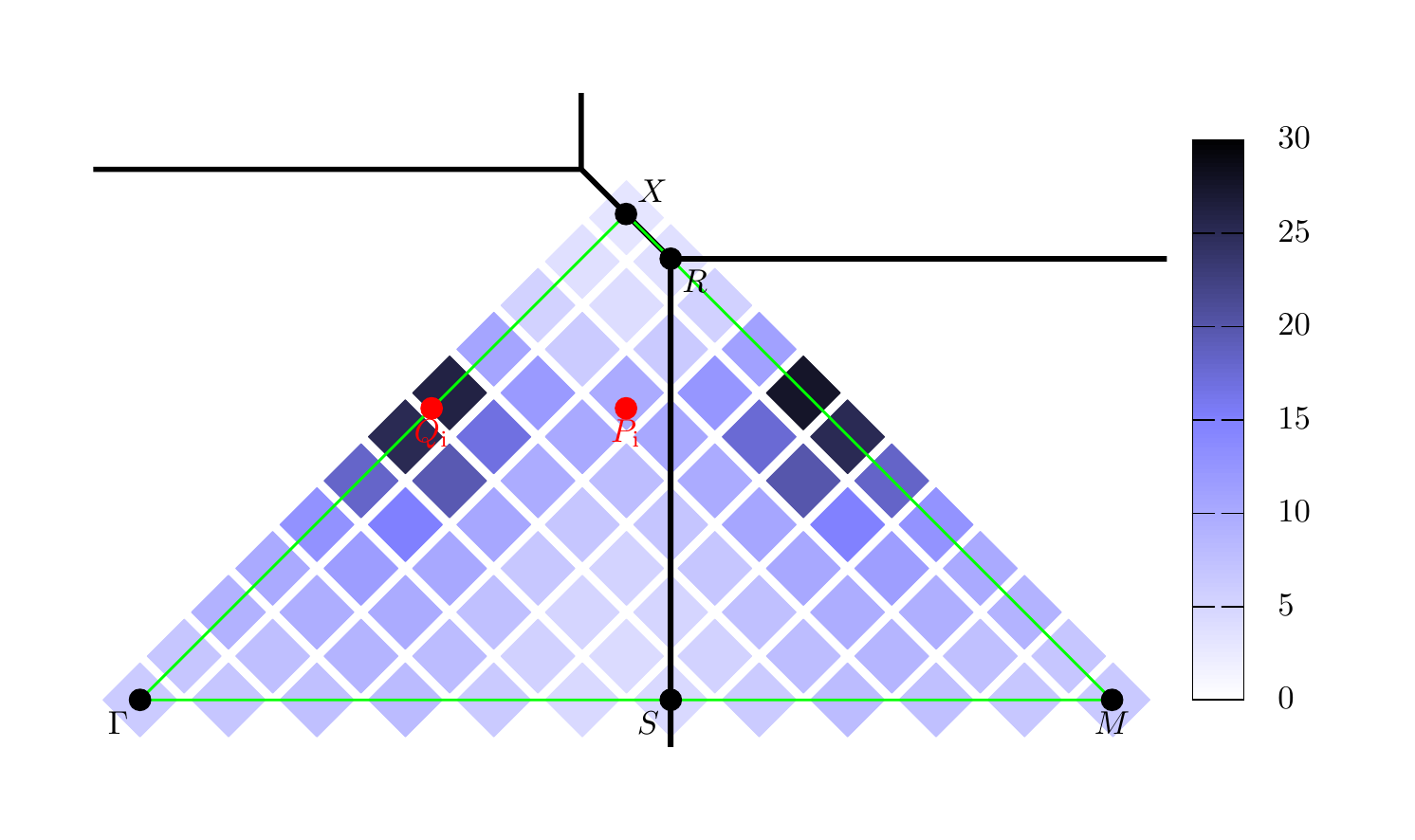}\\[-0.5cm]
(b)\hspace*{-0.25cm}\includegraphics*[width=8.5cm]{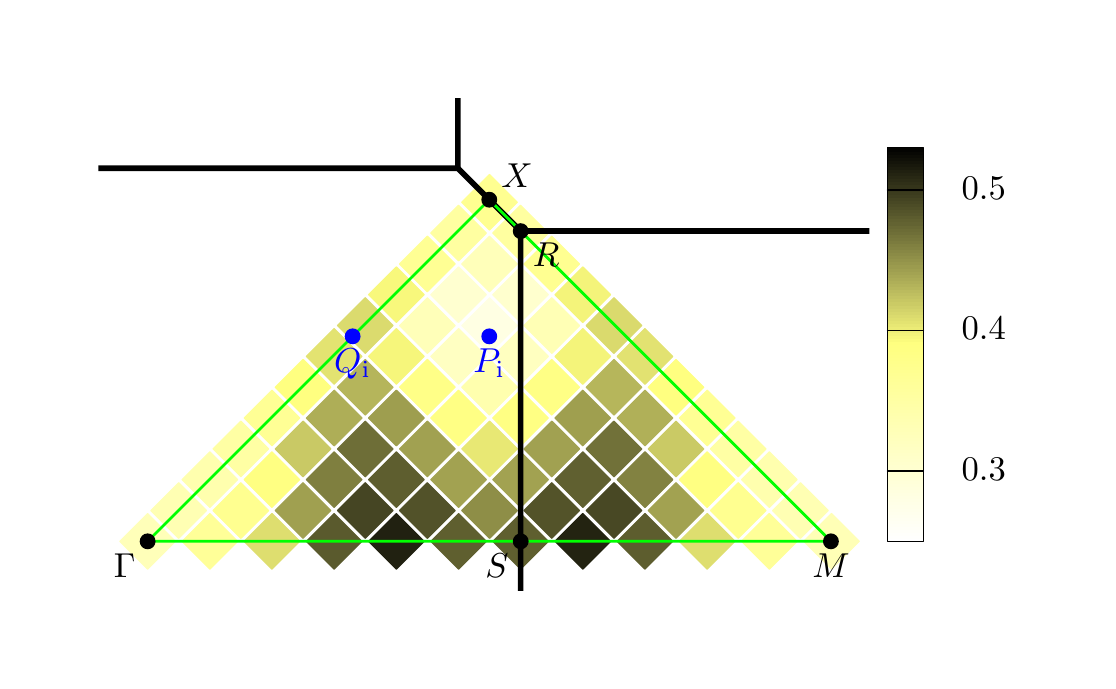}\\[-0.5cm]
(c)\hspace*{-0.25cm}\includegraphics*[width=8.5cm]{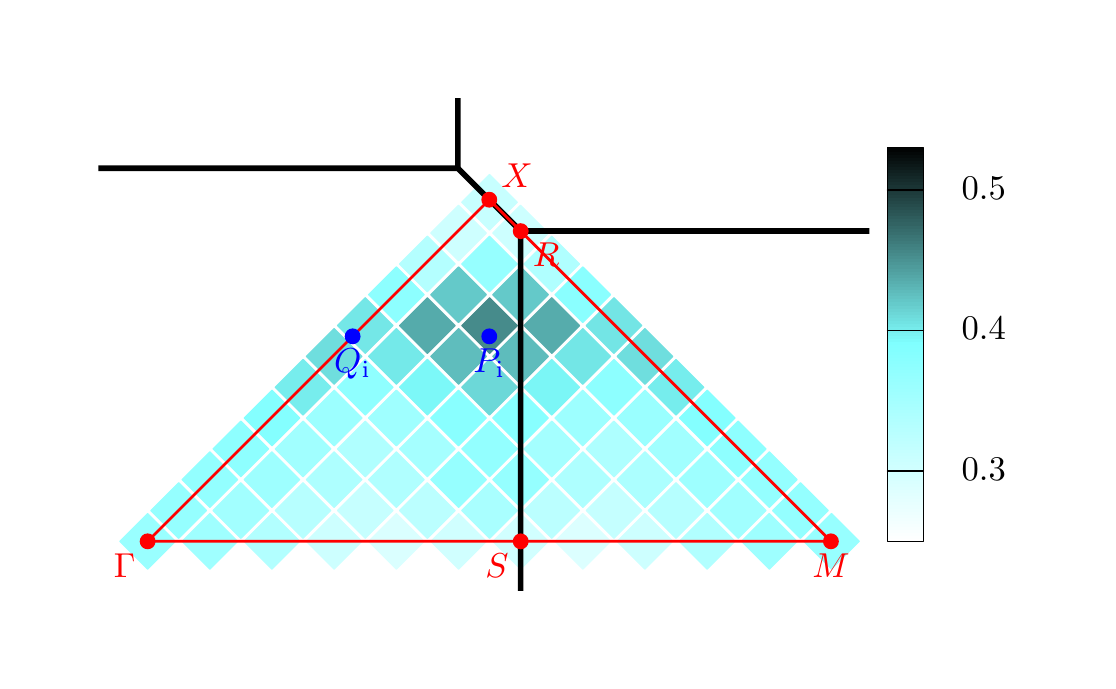}\\[-0.5cm]
(d)\hspace*{-0.25cm}\includegraphics*[width=8.5cm]{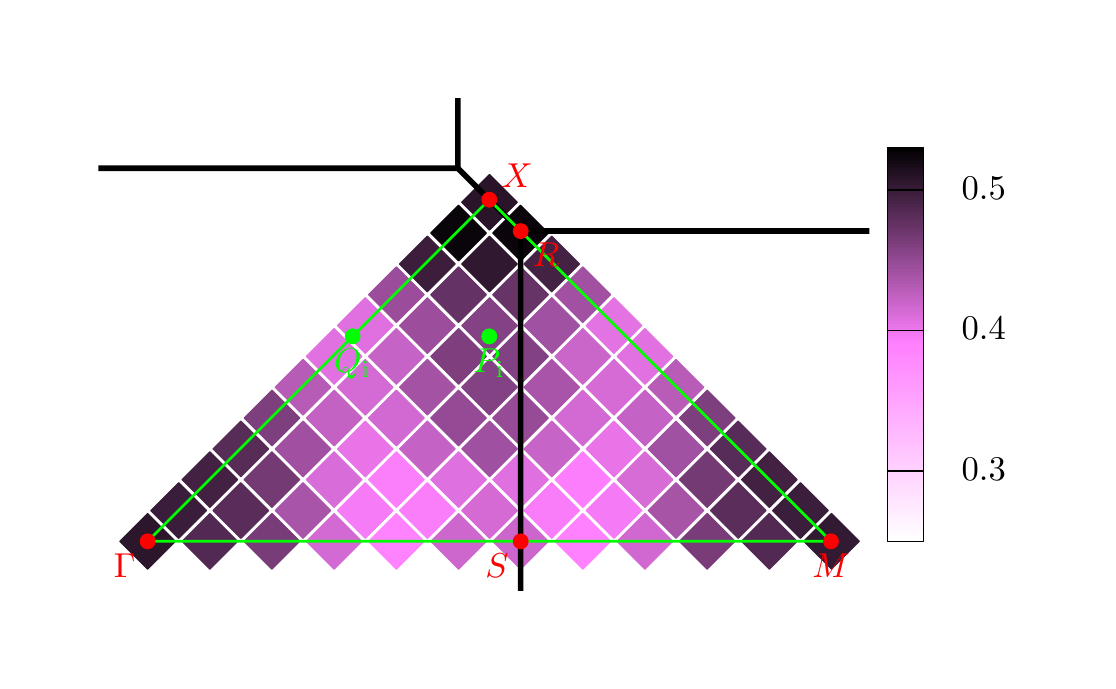}
\caption{(Color online) 
Data extracted from the orbital-resolved spin-susceptibility tensor at $T=193$\,K.
(a) Dominant eigenvalue of the susceptibility tensor throughot the irreducible wedge
of reciprocal space.
(b-d) $q$-dependent orbital contributions to the dominant eigenmode: (b) $v_{xz}$, 
(c) $v_{yz}$ and (d) $v_{xy}$.}\label{fig:sus-map}
\end{figure}

To identify the dominant particle-hole fluctuation in an unbiased way, the 
dominant eigenvalues of the susceptibility tensor $\chi^{\sigma\sigma'}_{mm'm''m'''}$ are 
monitored in the static limit $\omega=0$. The maximum eigenvalue in Sr$_2$RuO$_4$ 
corresponds to a mode that is diagonal in the orbital components throughout the
BZ. The eigenmode is indeed related to $S^{z}$ and exhibits the form 
$V^{\rm max}({\bf q})\sim v_{xz}({\bf q})S_{xz}^{z}+v_{yz}({\bf q})S_{yz}^{z}+v_{xy}({\bf q})S_{xy}^{z}$, 
where the normalization $\Sigma_{\kappa}v_{\kappa}^2({\bf q})=\nicefrac{1}{2}$ holds. 
Note that the
orbital diagonality is not trivial and there are other cases with a more complex 
structure~\cite{boe14}. The dispersion of this dominant eigenvalue, plotted in 
Fig.~\ref{fig:sus}b, shows a significant temperature dependence. Whereas the peak
close to ${\bf Q}_i$ appears already at rather high $T$ (with notable shifting in $q$-space
when lowering the temperature) a clear peak shoulder along $\Gamma X$ towards $\Gamma$
sets in only below room temperature. That shoulder is again in very good agreement with
experiment~\cite{bra02}. Directly at $\Gamma$ the eigenvalue grows with $T$, but a large
response pointing to an obvious FM contribution does not show up. 

A full 2D mapping of the
dominant-eigenvalue intensity within the irreducible wedge of the BZ readily
marks ${\bf Q}_i$ as the hot spot for spin-like fluctuations (see Fig.~\ref{fig:sus-map}a).
Most interesting in this context is the variation of the respective orbital contributions
$v_{xz}, v_{yz}, v_{xy}$ throughout the BZ, depicted in Fig.~\ref{fig:sus-map}b-d. For
instance, the eigenmode at $\Gamma$ is characterized by $v_{xz}({\bf 0})=v_{yz}({\bf 0})=0.34$ 
and $v_{xy}({\bf 0})=0.52$. Thus the $d_{xy}$ orbital makes the major contribution to
the spin fluctuations at $\Gamma$. Within the irreducible wedge, the regions of dominant
weight of the different $t_\mathrm{2g}$ orbitals are mutually exclusive. As expected, $d_{xz}$
and $d_{yz}$ contribute strongly along the original 1D-like Fermi-surface 
directions, while the $d_{xy}$ weight is largest around the high-symmetry points $\Gamma$,
$X$ and $M$. Quite surprisingly, right at the spin-fluctuation peak position 
${\bf Q}_i$ {\sl all} the orbitals have an {\sl equal} share, i.e. 
$v_{xz}({\bf Q}_i)=v_{yz}({\bf Q}_i)=v_{xy}({\bf Q}_i)=0.41$. Hence a straightforward 
$d_{xz}$-$d_{yz}$ nesting scenario cannot explain the ${\bf Q}_i$-peak. Spin
flucutations at the latter $q$-point have a manifest multi-orbital character. Our finding
is supported by the recent work of Arakawa~\cite{ara14}.

\section{Summary}
Our realistic many-body approach based on DFT+DMFT allows for an extended study of the 
multi-orbital electronic properties in the single-layer Ruddlesden-Popper compound 
Sr$_2$RuO$_4$. By means of generic lattic susceptibilities at strong coupling, i.e.
including vertex contributions, we examine the spin-fluctuation spectrum depending on 
wave vector, frequency and temperature. Good agreement with available INS data is obtained, 
the experimental findings concerning the relevant energy and temperature scales are confirmed. 

From an eigensystem analysis of the generic particle-hole susceptibility tensor it is indeed 
possible to designate a magnetic eigenmode proportional to $S^{z}$ as the dominant 
one. The orbital contributions to that mode vary over the Brillouin zone, describing some 
anisotropy in the spin response. Importantly, at the major peak position 
${\bf Q}_i=(0.3,0.3)$ in the ${\bf q}_z=0$ plane the latter has an intrinsic multi-orbital nature, 
involving fluctuations in all three $t_\mathrm{2g}$ orbitals equally, in contrast to a combination of 
1D-nesting pictures.
This finding is important not only for a deeper understanding of the magnetism in 
Sr$_2$RuO$_4$, but moreover for theoretically approaching the pairing mechanism at very
low temperatures. Therefore advancing the present scheme towards an assessment of the 
particle-particle susceptibility would be promising. Investigating the 
influence of spin-orbit coupling~\cite{ng00} is an additional important aspect.

\acknowledgments
The authors are indebted to M. Ferrero, A. Georges and O. Parcollet for helpful
original discussions on the vertex implementation.
The work benefited from financial support through the DFG-FOR1346. 
Calculations were performed at the North-German Supercomputing Alliance (HLRN)
under Grant No. hhp00035.

\bibliographystyle{eplbib}
\bibliography{bibextra}

\end{document}